\documentclass[]{aa}
\usepackage{graphicx}
\usepackage{txfonts}
\usepackage{natbib}
\bibpunct{(}{)}{;}{a}{}{,}

\begin{document}

\title{Diffusive shock acceleration in radiation dominated environments}

\author{G. Vannoni\inst{1}\thanks{IMPRS fellow - \email{Giulia.Vannoni@mpi-hd.mpg.de}} \and  S. Gabici\inst{2} \and F. A. Aharonian\inst{2,1}} 

\institute{Max-Planck-Institut f\"ur Kernphysik, Saupfercheckweg 1, Heidelberg 69117, Germany; 
\and Dublin Institute for Advanced Studies, 31 Fitzwilliam Place, Dublin 2, Ireland.}

\abstract
{Radio, X-ray, and gamma-ray observations provide us with strong evidence of particle acceleration to multi-TeV energies in various astrophysical sources. Diffusive shock acceleration is one of the most successful models explaining the presence of such high-energy particles.}
{We discuss the impact of inverse Compton losses on the 
shock acceleration of electrons that takes place in radiation dominated 
environments, i.e. in regions where the radiation energy density exceeds that 
of the magnetic field.}
{We perform a numerical calculation, including an energy-loss term in the transport equation of accelerated particles.}
{We discuss the implications of this effect on the hard X-ray 
synchrotron and gamma-ray inverse Compton radiation, produced by shock-accelerated electrons in young supernova remnants in the presence of large 
radiation fields (e.g. in the Galactic centre). We also discuss possible 
implications of our results for clusters of galaxies and gamma-ray binaries.}
{We demonstrate that the inverse Compton losses of electrons, in the 
Klein-Nishina regime, lead to spectra of ultra-relativistic electrons that may significantly differ from classical diffusive shock acceleration solution. The most prominent feature is the appearance of a pile-up in the spectrum around the cut-off energy.}

\maketitle

\section{Introduction}
\label{Introduction}

Diffusive shock acceleration (DSA) is the most successful and widely accepted mechanism for explaining the acceleration of cosmic rays in many astrophysical environments.
Despite the basic physics of DSA being very robust and reasonably well understood (see \citealp{bl_eich} and \citealp{malk} for a review), some crucial issues still need to be addressed.

In its simplest, test-particle approach the theory of DSA predicts an energy spectrum for the accelerated particles, which is a featureless power law $f(p) \propto p^{-\alpha}$ up to a maximum energy where the spectrum cuts off. For strong, nonrelativistic shocks, where the mechanism is believed to operate very efficiently, the slope of the spectrum below the cut-off converges to the canonical slope $\alpha = 4$, independently of the details of the acceleration.

The value of the maximum energy attained by particles is determined by the competition between the acceleration time and the shortest of the three time scales: {\textit i)} particle escape time from the accelerator, {\textit {ii})} particle energy loss time, and {\textit {iii})} age of the accelerator.
Therefore, while the energy spectrum below the cut-off is almost universal, the spectral shape of the cut-off depends dramatically on both the details of the acceleration mechanism and the physical process determining the particle maximum energy.
In the case of electrons, which we consider in this paper, the maximum energy is in most cases limited by radiative synchrotron and inverse Compton losses in the ambient magnetic and photon fields.

The missing piece of information in the model is the diffusion coefficient of particles close to the shock, which determines the acceleration rate. For a given diffusion coefficient, the problem is well-defined and can be solved once the value of the magnetic field and the spectrum of the ambient radiation are specified.
Although there is some uncertainty about this coefficient, one can reasonably assume that, due to the high level of turbulence expected close to the shock, the mean free path of particles is close to their Larmor radius $r_L$. Under these circumstances, diffusion proceeds close to the slowest possible (so-called Bohm) rate with diffusion coefficient $D \sim r_L c /3$.

Several approaches can be found in the literature to the problem of electron acceleration at shocks in the presence of radiative losses. \citet{bul}, \citet{webb}, and \citet{M} considered the case of a diffusion coefficient constant in momentum and space.
Recently \citet{zi} numerically solved the problem for $D(p)$ with an arbitrary momentum dependence.
All these approaches are limited to the case of dominant synchrotron losses.

In this paper, we numerically solve the problem for a general form of both the diffusion coefficient and the energy-loss rate.
In particular, we focus on the case, never considered before, in which the accelerator is embedded in a strong radiation field, characterised by a much higher energy density than the magnetic one.
Under this circumstance, inverse Compton losses dominate synchrotron losses and, at the highest energies, modifications to the electron spectrum are expected with respect to the case of synchrotron/Thomson losses, due to the transition between Thomson and Klein-Nishina regimes.
This, in turn, strongly affects the spectrum of the radiation emitted by accelerated particles.
We calculate the exact shape of the spectrum in the whole energy range and show that, in the case of Klein-Nishina losses, the particle distribution at the shock has a broad cut-off, due to the shallow energy dependence of the loss rate.
Another important effect of Klein-Nishina losses is to harden the downstream electron spectrum close to the maximum energy, leading to the formation of a pronounced pile-up. Such a feature can be observed in the photon spectrum, in particular for the synchrotron emission. The effect on the inverse Compton emission is less pronounced because the Klein-Nishina cross-section intervenes twice, in opposite directions, to harden the electron spectrum and to soften the photon one leading to an almost exact compensation.

We describe the details of our model in Sect. \ref{The Model} and discuss the main characteristics of the calculation by applying it to an ideal example in Sect. \ref{Results}. In Sect. \ref{Applications} we then apply our results to several astrophysical environments where the radiation field-energy density dominates, namely to supernova remnants located close to the Galactic centre, shocks in binary systems and accretion shocks around massive clusters of galaxies.
For these systems we evaluate the spectrum of the accelerated particles, together with the spectra of both the emitted synchrotron and inverse Compton radiation.
We summarise in Sect. \ref{Conclusions}. The presence of a strong pile-up at the high-energy end of the synchrotron spectrum is the most remarkable feature.

\section{The model}
\label{The Model}

We consider a plane-parallel shock where the fluid moves (in the shock rest-frame) along the \textit{x}-axis from $- \infty$ far upstream, to $+ \infty$ far downstream, and the shock is located at $x=0$.
We assume the velocity of the shock to be nonrelativistic.

The transport equation for the particles distribution function, including the presence of energy losses, reads as

$$\frac{\partial f(x,p,t)}{\partial t}+u\frac{\partial f(x,p,t)}{\partial x}-\frac{\partial }{\partial x}\left( D(x,p)\frac{\partial f(x,p,t)}{\partial x}\right)-\frac{p}{3}\frac{\partial u}{\partial x}\frac{\partial f(x,p,t)}{\partial p}$$
\begin{equation}
-\frac{1}{p^2}\frac{\partial }{\partial p}(p^2 L(x,p) f(x,p,t))=Q(x,p),
\label{trans}
\end{equation}
where $L(x,p)=-\dot p$ is the loss rate taken to be positive and $Q(x,p)$ is the injection term; $u$ represents the bulk velocity of the plasma in the shock rest-frame.
In the following we assume that the diffusion coefficient in the up and downstream regions $D_{1,2}$ does not depend on $x$ and has the functional form $D(p)=D_0 p^\beta$ (although any expression for $D_{1,2}(x,p)$ can be easily implemented in the code) and that injection happens at the shock surface as a delta function in energy:
$$Q(x,p)=Q_0\delta(x) \delta(p-p_0).$$

The formulation is general so that it can be applied to both protons or electrons, plugging in the relevant energy-loss channels, namely proton-proton and proton-gamma interactions, for the hadronic channel, and synchrotron and inverse Compton (IC) emission for the leptonic one.
In this paper we study the acceleration of electrons undergoing synchrotron and inverse Compton losses, in the case where the IC is dominant. This is the case if acceleration takes place in a region where the energy density of the ambient radiation field exceeds the magnetic one: $U_{rad}\gg B^2/(8\pi)$. In particular we are interested in studying the effect produced on both the electron and the photon distributions when IC losses proceed in the Klein-Nishina regime, namely when $\epsilon E/(mc^2)^2\ge 1$ with $E$ being the electron energy, $m$ the electron mass, and $\epsilon$ the ambient photon's energy.

In the following we assume that the background radiation field is isotropic and that the background magnetic field is constant in the upstream and downstream regions and the two are related by $B_2=\xi B_1$, where $\xi$ is the compression factor. Such a parameter varies between two values: $\xi=1$ for a pure parallel shock and $\xi=4$ for a perpendicular shock.


\subsection{Electron spectra}

We solve Eq. (\ref{trans}) using a finite difference numerical scheme, implicit in time to guarantee unconditional stability and explicit in momentum.
To solve the problem we need to insert the boundary conditions at the shock and at upstream/downstream infinity.
For the boundary at the shock we consider again Eq. (\ref{trans}) and integrate it between $0_{-}$ immediately upstream and $0_{+}$ immediately downstream, obtaining

\begin{equation}
\frac{1}{3}(u_1-u_2)p\frac{\partial f_0}{\partial p}=D_2 \frac{\partial f_0}{\partial x}\Big|_2 - D_1 \frac{\partial f_0}{\partial x}\Big|_1+Q_0 \delta(p-p_0)
\label{boundary}
\end{equation}
where the subscript $1$ refers to the upstream region and $2$ to the downstream one. At $\pm \infty$ we set $f(x,p)=0$. 

It is convenient to introduce two dimensionless variables in place of $p$ and $x$. 
The first variable is $p/p^*$, where $p^*$ is a parameter that estimates the cut-off momentum for the electron distribution, evaluated by imposing equilibrium between the momentum gain in one acceleration cycle,
\begin{equation}
\Delta p_{acc}=\frac{4p}{3}\frac{(u_1-u_2)}{c},
\end{equation}
and the momentum loss per cycle,
\begin{equation}
\Delta p_{loss}=L_1(p)\Delta t_1+L_2(p)\Delta t_2,
\end{equation}
where the mean residence time in the up and downstream regions is $\Delta t_{1,2}=4 D_{1,2}(p)/cu_{1,2}$ (see \citealp{webb}).

For the spatial coordinate we operate a change of variable both up and downstream. To reduce the wide range in $|x|$ from 0 to $\infty$ to one more numerically feasible, we use an exponential variable:

\begin{equation}
z=exp\left[-\frac{|x|}{x_0}\right],
\end{equation}
where $x_0$ is a characteristic length scale. The new variable now conveniently ranges from 0 to 1.
In the upstream region, except for energies close to the cut-off, the accelerated particles can propagate one diffusion length ahead of the shock.
The situation is different downstream, where the maximum distance from the shock a particle can reach, in steady state, is determined by energy losses.
Thus, a reasonable choice for the length scale $x_0$ is
\begin{equation}
x_{0,1}=\frac{D_1(p)}{u_1}
\end{equation}
and
\begin{equation}
x_{0,2}=u_2\tau_L,
\end{equation}
for upstream and downstream, respectively. The energy loss time is given by $\tau_L=p/L(p)$.


Once the solution $f(x,p)$ of Eq. (\ref{trans}) is found, we evaluate the total spectrum integrated over space, $F(p) = \int f(x,p)dx$, which is the quantity required to evaluate the radiation spectrum emitted in the shock region. 
It is worth mentioning that the steady state integrated spectrum, below the cut-off energy, follows a power law. 
In the downstream region, in case of Thomson cooling, the differential spectrum in momentum space has index $\delta=\alpha+1=5$. Upstream, the maximum distance from the shock a particle can travel is essentially its diffusion length; therefore, for Bohm-like diffusion, this results in $\delta=\alpha-1=3$.

\subsection{Photon spectra}

Once the electron spectrum is obtained, we can calculate the spectra of the radiation emitted both via synchrotron and IC.
For synchrotron radiation the energy flux at an energy $\epsilon$ is given by
\begin{equation}
\Phi(\epsilon)=\frac{\sqrt{3}Be^3}{hmc^2}\int p^2 dp F(p) K(\epsilon/\epsilon_c),
\end{equation}
where $\epsilon_c=h\nu_c$ is the energy corresponding to the critical frequency $\nu_c=3eBp^2/(4\pi m^3c^3)$ and $K(\epsilon/\epsilon_c)$ is the emission produced by the single electron of momentum $p$, charge $e$, and mass $m$; $h$ represents the Planck constant.

In our case $F(p)$ is the total electron spectrum integrated up and downstream. The exact expression for the kernel function $K(\epsilon/\epsilon_c)$ in the case of a turbulent magnetic field was derived in \citet{kernel}. With several percent accuracy, this can be approximated by the analytical expression obtained in \citet{zi}:
\begin{equation}
K(\epsilon/\epsilon_c)=\frac{1.81 e^{-\epsilon/\epsilon_c}}{\sqrt{(\epsilon/\epsilon_c)^{-2/3}+(3.62/\pi)^2}}.
\end{equation}
We adopt this simpler expression.

The inverse Compton energy flux for an isotropic distribution of soft photons $n(\epsilon')$ upscattered by a population of electrons with spectrum $F(p)$ is \citep{bg}:
$$\Phi(\epsilon)= \frac{2 \pi e^4 \epsilon}{c}\int dp F(p)\int \frac{n(\epsilon') d\epsilon'}{\epsilon'}$$
\begin{equation}
\left[2q {\rm ln} q+(1+2q)(1-q)+\frac{1}{2}\frac{\epsilon ^2}{pc(pc-\epsilon)}(1-q) \right],
\end{equation}
where
$$q=\frac{\epsilon}{\frac{4\epsilon' pc}{(mc^2)^2}(pc-\epsilon)}.$$

For simplicity, in the following we assume that the background radiation field is a black body, though these calculations may easily be extended to any background radiation spectrum. 
Depending on the environment we want to model, we consider either a single or a superposition of multiple Planck distributions, introducing a dilution factor $\eta$ for each:
\begin{equation}
n(\epsilon')=\eta\frac{1}{\pi^2 \hbar^3 c^3}\frac{\epsilon'^2}{e^{\epsilon'/kT}-1},
\end{equation}
where $k$ is the Boltzmann constant and $T$ the black body temperature. 

\section{Results}
\label{Results}

We present here the results of our calculation by means of an ideal case to illustrate the features and characteristics of the problem we are studying.
We consider a nonrelativistic shock expanding in a medium where the magnetic field value is $B_1=10~\mu$G. For such a field, the magnetic energy density is $B^2/8\pi = 2.5$ eV/cm$^3$.
To keep a limited number of model parameters, in this example we assume the same value for the magnetic field upstream and downstream (i.e. $\xi=1$, strictly valid for a perfectly parallel shock). In the next section, when we apply our results to astrophysical objects, we include the compression of the magnetic field at the shock.
We assume that the accelerator is embedded in an isotropic background radiation field. We further assume a diluted black body radiation spectrum (i.e. with a reduced energy density) at a temperature of $kT=0.3~\rm{eV}$ (i.e. in the near infrared energy band), corresponding to $T=3500~\rm{K}$. We then assume the radiation energy density to be $U_{rad}=5\times 10^4~\rm{eV/cm^3}$, a factor $2\times 10^4$ greater than the magnetic one. (As we point out in the next section, this high value is similar to the one measured in the $\le 1~\rm{pc}$ Galactic centre region \citep{dav}).
\begin{figure}[]
\resizebox{\hsize}{!}{\includegraphics{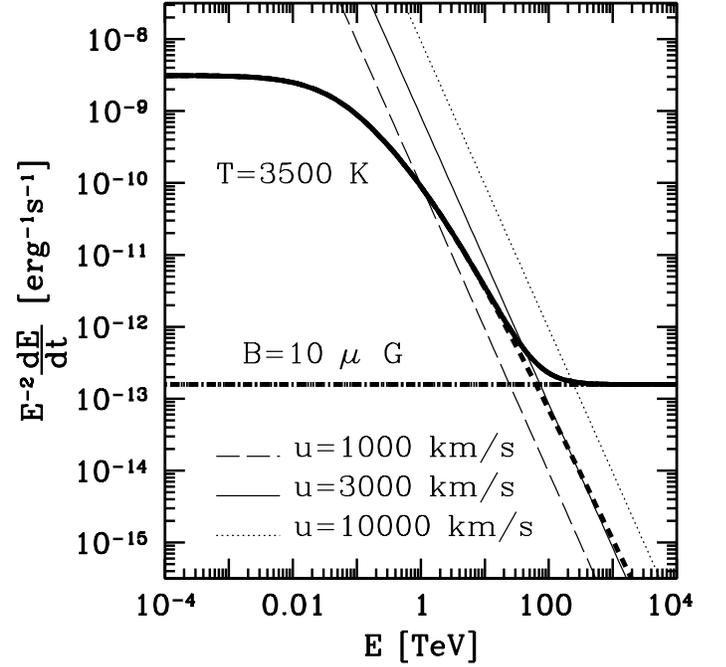}}
\caption{Energy-loss rates and shock acceleration rates for a magnetic field $B=10~\rm{\mu G}$ and a diluted black body radiation field at a temperature $T=3500~\rm{K}$. All curves have been multiplied by $E^{-2}$ so that Thomson losses correspond to horizontal lines. Thick lines: synchrotron loss rate (dot-dashed), inverse Compton loss rate (dashed) and the sum of the two (solid). Thin curves: acceleration rates for $u_1=1000~\rm{km/s}$ (dashed), $u_1=3000~\rm{km/s}$ (solid) and $u_1=10000~\rm{km/s}$ (dotted).}
\label{losEx}
\end{figure}

In Fig. \ref{losEx} we plot the energy-loss rates due to synchrotron losses in a magnetic field of $10~\rm{\mu G}$ and inverse Compton scattering off a radiation field at $3500~\rm{K}$, as a function of the particle energy. The sum of the two is also represented, as well as the acceleration rates for DSA in the same magnetic field and for three different values of the shock velocity ($u_1=1000~\rm{km/s}$, $u_1=3000~\rm{km/s}$, and $u_1=10000~\rm{km/s}$). In this figure all curves have been multiplied by $E^{-2}$ so that Thomson losses correspond to horizontal lines.
The intersection between the acceleration rate curve and the total energy loss rate indicates the point in the particle energy spectrum where the acceleration due to DSA is compensated by radiative losses so the cut-off sets in.

Figure \ref{losEx} shows how the energy dependence of energy losses changes behaviour.
At low energies ($E \ll (mc^2)^2/\epsilon_{ph}$), inverse Compton proceeds in the Thomson regime ($dE/dt \propto E^2$), but at high energies the losses enter the Klein-Nishina regime and the dependence of the process on energy changes ($dE/dt \sim \textrm{ln} E$). At even higher energies synchrotron losses become dominant and an $E^2$ law for the loss rate is recovered.



\begin{figure}[]
\includegraphics[width=0.4\textwidth]{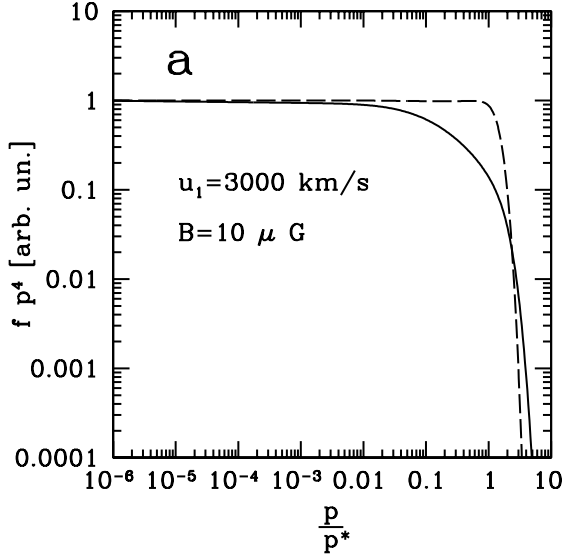}\\
\\
\includegraphics[width=0.4\textwidth]{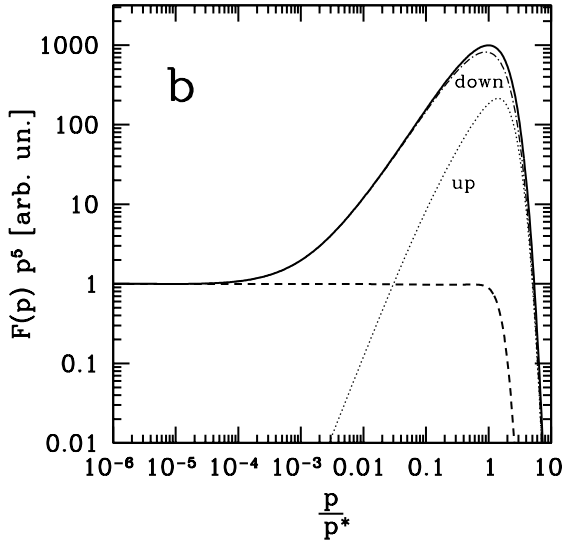}
\caption{\textbf{a)} Electron spectrum at the shock location in the case of inverse Compton dominated losses (solid line) and synchrotron cooling dominated case (dashed line) for a magnetic field $B=10~\rm{\mu G}$ equal up and downstream, shock velocity $u_1=3000~\rm{km/s}$ and compression ratio $R=4$. The background radiation field is assumed to be a diluted black body at $T=3500~\rm{K}$ with energy density $U_{rad}=5\times 10^4~\rm{eV/cm^3}$.
The momentum scale for each curve is normalised to the parameter $p^*$. The normalisation on the \textit{y}-axis is in arbitrary units.
\textbf{b)} Electron distributions integrated over space. In the case of the IC dominated case (solid line), we also show separately the two components upstream (dotted) and downstream (dash-dotted). The dashed line represents the synchrotron dominated case.}
\label{electEx}
\end{figure}

We solve the transport equation for the accelerated electrons choosing the value of $3000~\rm{km/s}$ for the shock velocity. For such a choice of parameters, losses proceed deep in the KN regime. The shock is assumed to be strong, i.e. compression ratio $R=4$, and we also assume a Bohm-type diffusion coefficient, namely $D=D_0p^{\beta}$ with $\beta=1$ and $D_0=c^2/(3 e B)$. In this example we make the assumption that the radiative energy loss time scale is much shorter than any other loss time scale in the system that could limit the acceleration, such as escape or the finite lifetime of the accelerator,
at all the energies we consider. Therefore we look for the steady state spectrum.
The resulting electron energy spectrum at the shock location is shown in Fig. \ref{electEx}a,  
\begin{figure}[]
\includegraphics[width=0.4\textwidth]{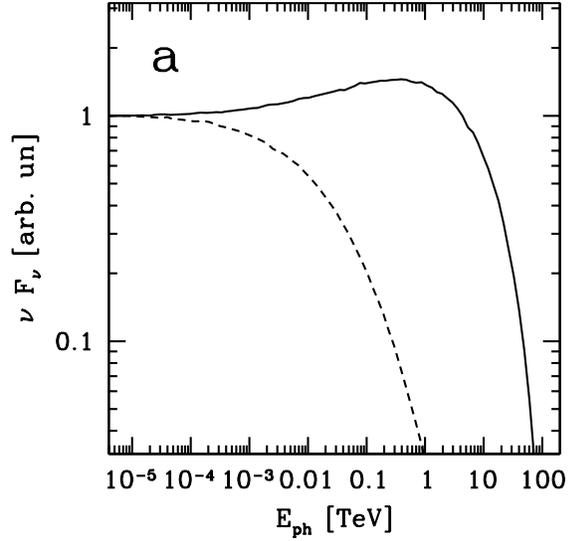}\\
\\
\includegraphics[width=0.4\textwidth]{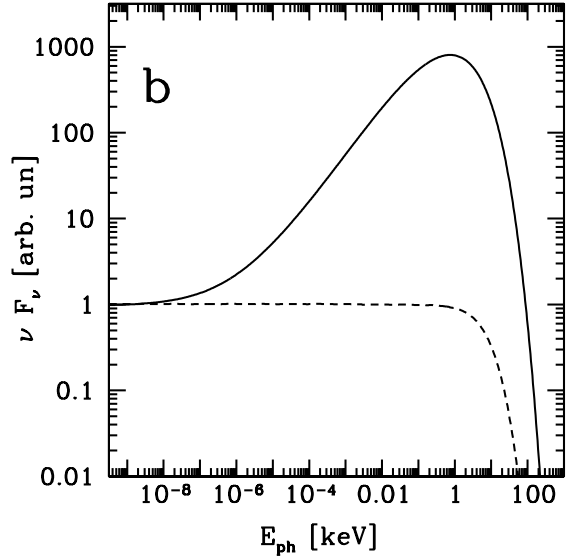}
\caption{\textbf{a)} Inverse Compton photon distribution produced by the electron spectra in Fig. \ref{electEx}b on a diluted black body at $T=3500~\rm{K}$ for the case of dominant IC losses (solid line) and for the case of dominant synchrotron cooling (dashed line).
\textbf{b)} Synchrotron emission from the two electron distributions in Fig. \ref{electEx}b, the magnetic field value is $10~\rm{\mu G}$.}
\label{radEx}
\end{figure}
together with the spectrum expected for the same values of the parameters but suppressing the radiation energy density (i.e. when synchrotron losses dominate).
For an easier comparison we rescaled the momentum scale for each curve to its cut-off value $p^*$. Of course the actual value of $p^*$ is different in the two cases. The normalisation on the \textit{y}-axis is in arbitrary units. The shape of the spectrum is modified by the decrease of the IC cross-section at high energies: the cut-off region is broader and the decay is shallower, compared to the synchrotron case.
The result may be understood most easily from the rates in Fig. \ref{losEx}, which highlights the weak dependence on energy of Klein-Nishina losses.

To obtain the overall radiation produced by this population of electrons we have to consider the spatially integrated spectra $F_1=\int_{-\infty}^0 f(x,p)dx$ and $F_2=\int_0^{\infty} f(x,p)dx$. Figure \ref{electEx}b shows the total spectrum $F(p)=F_1(p)+F_2(p)$ resulting from the integration
for both cases of IC dominant losses and of synchrotron dominant losses.
As can be seen, the most pronounced feature appears in the integrated spectrum: the softer dependence on energy of the energy loss rate in the KN regime, as compared to the Thomson regime, causes a significant pile-up around the cut-off energy. For this case, we also plotted the upstream and downstream distributions separately to show that the major contribution to the total spectrum is provided by the downstream distribution. In this region all particles are eventually advected away from the shock with the same velocity $u_2$. They also diffuse to a distance that depends on their energy and on time as $x_D=\sqrt{4 D(p) t}$. For particles with energy below the cut-off energy, advection is always dominant and the maximum distance to which they can propagate is determined by their loss time scale and is $x_{ad}=u_2 \tau_L$. The diffusion length at the same time is given by $x_D=\sqrt{4 D(p)\tau_L}$. Both the length scales increase with $\tau_L$. In particular, the two are comparable at the cut-off energy where $\tau_L=t_{acc}\simeq D/u_2^2$. At lower energies diffusion is negligible and advection sets the maximum propagation distance.
In the first approximation, the particle integrated spectrum downstream can be written as $F_2(p)=f_0 x_L$, with $x_L=\textrm{min}\{x_{ad}, x_D\}$.
For Thomson scattering $\tau_L\propto E^{-1}$, determining that the higher the particle energy, the smaller the distance it can travel away from the shock before losing its energy. For this reason the integrated spectrum in momentum space downstream, at energies below the cut-off where $x_L=x_{ad}$, is $F_2(p)= f_0 u_2 \tau_L\propto p^{-\alpha-1}$. In the case of KN losses we have an opposite trend: $\tau_L\propto E/\textrm{ln} E$, so that high-energy particles can propagate farther than low-energy ones, and the resulting spectrum is harder. This effect produces the pile-up when integrating the spectrum over space. In the case we are considering in this section, we assume that the loss time scale is the limiting one at every energy in the range we consider. In real physical environments this may not be the case. At low energies, where the Thomson regime is recovered also in the IC case, the time scale of energy losses increases with decreasing energy, therefore it could at some point become more than the age of the system ($\tau_{age}$) itself. The maximum energy of the accelerated particles is still determined by the energy losses, but the low-energy part of the spectrum does not have enough time to cool and the spectrum results in $F_{2}=f_0u_2\tau_{age}\propto p^{-\alpha}$.

With the total spatially integrated spectrum obtained, we calculate the radiation emitted. Our results are shown in Fig. \ref{radEx}.
Panel \textbf{a} shows the inverse Compton radiation resulting from the upscattering of the background radiation, when losses are Compton-dominated, compared to the dominant synchrotron losses case. 
The spectrum pile-up is not so remarkable in the first case because the effect of the Klein-Nishina cross-section acts twice: while hardening the electron's distribution, it softens the IC photons' one, so that the two effects almost cancel each other out.

On the other hand, the accelerated particles' features have a sharp imprint on the synchrotron spectrum (Fig. \ref{radEx}b), which carries direct information on the electron distribution.
Compared to the case of pure synchrotron cooling, the emission due to an electron distribution shaped by IC losses presents a pile-up of three orders of magnitude around the cut-off, which appears at keV energies for the chosen parameters.
This characteristic is a remarkable signature to distinguish the two different scenarios for electron radiative losses.

\section{Applications}
\label{Applications}

The results we obtained in the previous section may have broad astrophysical applications. To demonstrate this, we apply the results of our calculations to three specific astrophysical environments where strong nonrelativistic shocks may form and in which the radiation energy density may dominate over the magnetic field energy density.
The three cases we consider are: {\textit {i})} a supernova remnant in the Galactic centre region, {\textit {ii})} the accretion shock surrounding clusters of galaxies and {\textit {iii})} a shock in a microquasar jet.
For each of the considered cases, we evaluate both the electron spectrum and the spectrum of the emitted radiation.

\subsection{SNR in the Galactic centre}
\label{GC}

\begin{figure}[]
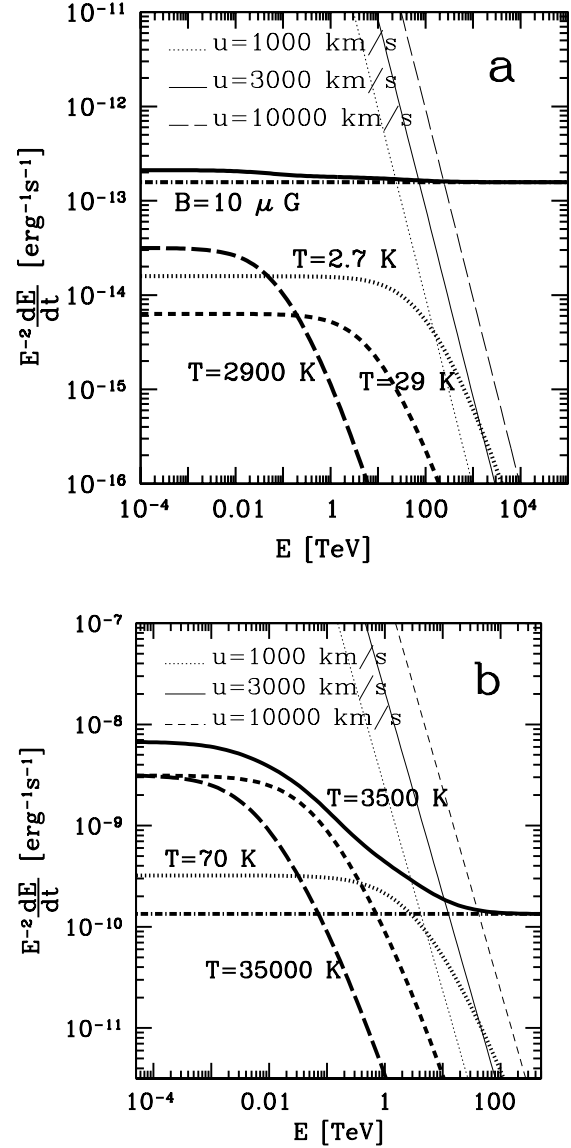

\includegraphics[width=0.4\textwidth]{9744fig6.epsi}\\
\\
\includegraphics[width=0.4\textwidth]{9744fig7.epsi}
\caption{\textbf{a)} Energy loss rates and acceleration rates, multiplied by $E^{-2}$, as a function of the electron energy $E$, for a supernova remnant in the Galaxy. The synchrotron rate (thick dot-dashed line) corresponds to a magnetic field $B=10~\rm{\mu G}$, the IC curves are calculated for three black bodies at temperature $T=2900~\rm{K}$ (optical), thick long-dashed line; $T=29~\rm{K}$ (far infrared), thick short-dashed line; $T=2.7~\rm{K}$ (CMB), thick dotted line. The thick solid curve represents the sum of the four loss contributions. Thin lines: acceleration rates for $B=10~\rm{\mu G}$ and shock velocity of $1000~\rm{km/s}$ (dotted), $3000~\rm{km/s}$ (solid), and $10000~\rm{km/s}$ (dashed).
\textbf{b)} Same as in panel \textbf{a} but for a magnetic field $B_1=100~\rm{\mu G}$ upstream and $B_2=4B_1$ downstream, and three black bodies at $T=35000~\rm{K}$ (UV/Optical), $T=3500~\rm{K}$ (NIR), and $T=70~\rm{K}$ (FIR).}
\label{lossSNR}
\end{figure}
\begin{figure}[]
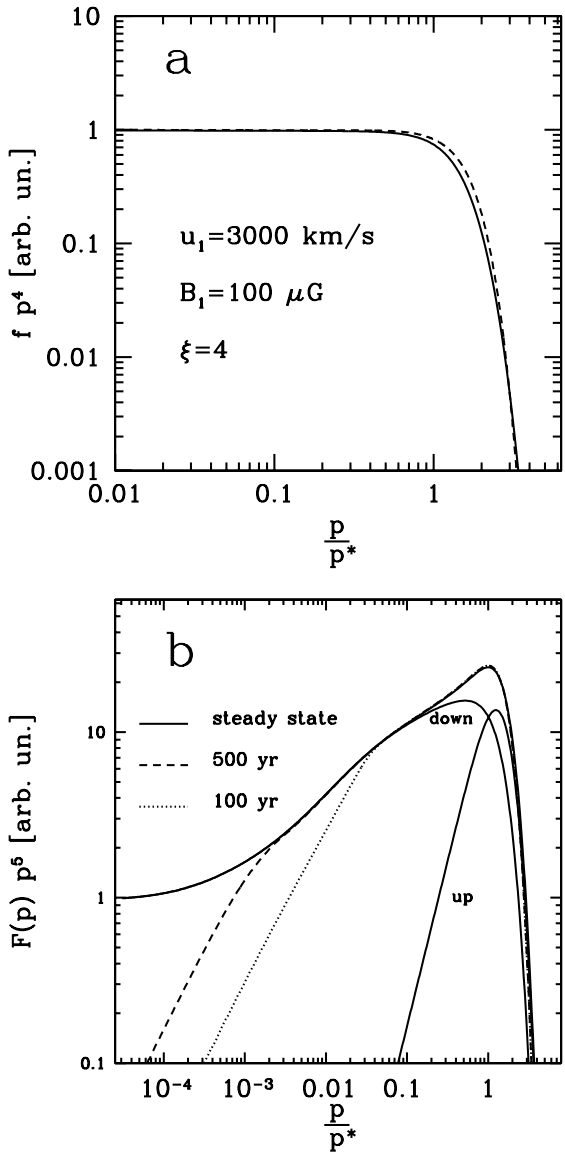

\includegraphics[width=0.4\textwidth]{9744fig8.epsi}\\
\\
\includegraphics[width=0.4\textwidth]{9744fig9.epsi}
\caption{\textbf{a)} Electron spectrum at the shock for a velocity of $3000~\rm{km/s}$ and a magnetic field $B_1=100~\rm{\mu G}$. The solid line is obtained in the case of dominant IC losses with the same parameters as in Fig. \ref{lossSNR}b; the dashed line corresponds to the case when the radiation energy density is suppressed and synchrotron cooling dominates.
\textbf{b)} Integrated spectra for the same values as the parameters of panel \textbf{a}. The three curves correspond to different maximum times of acceleration, as reported.}
\label{electGC}
\end{figure}
\begin{figure}[]
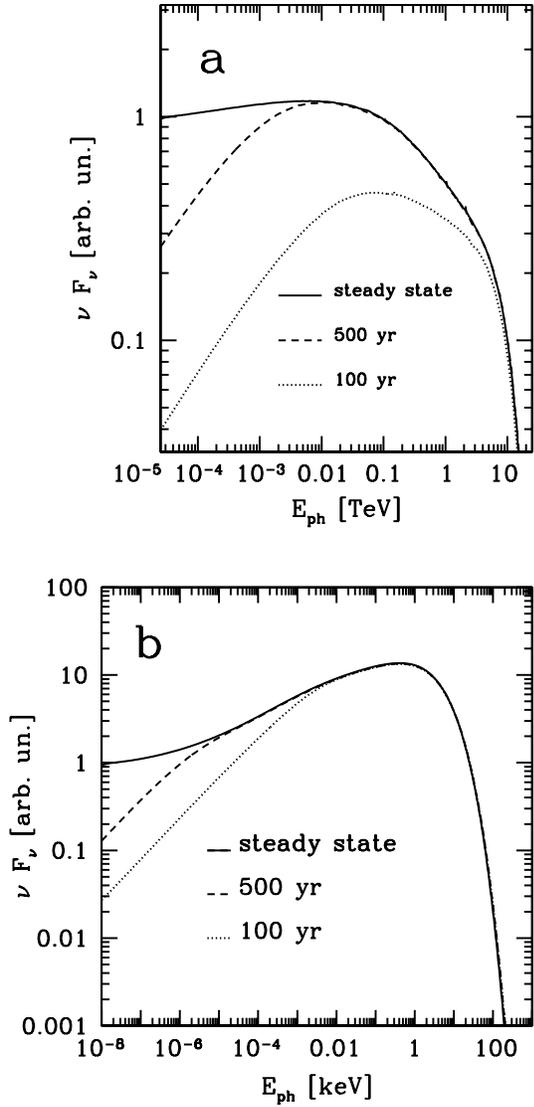

\includegraphics[width=0.4\textwidth]{9744fig10.epsi}\\
\\
\includegraphics[width=0.4\textwidth]{9744fig11.epsi}
\caption{Radiation emitted by the electron distributions in Fig. \ref{electGC}b. The up and downstream components have been summed up. \textbf{a)} Total inverse Compton spectra. The contributions of the three black bodies considered have been summed up.
\textbf{b)} Synchrotron radiation spectra.}
\label{radGC}
\end{figure}
Several observations in the radio and X-ray bands have confirmed supernova remnants (SNR) shocks as powerful accelerators of electrons up to tens of TeV \citep{bamba}.
We first consider a SNR in the galactic disc. The radiation field in the Galaxy consists of three distinct components: the optical/near-infrared (NIR), the far infrared (FIR) and the cosmic microwave background (CMB) radiation, with an energy density of $\sim 0.5, \sim 0.1, \sim 0.25 ~\rm{eV/cm^3}$, respectively \citep{mathis}.
In our calculation we evaluated an effective temperature for these fields from their peak energy and then assumed they can be approximated with a diluted black body distribution.
The Galactic magnetic field has a measured value of $\sim 3~\rm{\mu G}$ (see \citealt{wid} for a review), which corresponds to an energy density of $\sim 0.23~\rm{eV/cm^3}$. 
This implies that the average magnetic-field energy density is comparable to the radiation one. Moreover, the magnetic field can be significantly amplified in the presence of a shock that is efficiently accelerating particles \citep{ampl}, so that, in the acceleration region, $U_{mag}>> U_{rad}$ results. Therefore synchrotron losses dominate the inverse Compton (Fig. \ref{lossSNR}a).
In this case the contribution of the IC to the energy-loss rate is negligible and the cut-off energy and shape are determined by synchrotron cooling. 

Things are dramatically changed, though, if we consider an SNR in the Galactic centre region, where both the magnetic and the radiation fields are much stronger than the average Galactic values. 
Analogous to what is done in \citet{jim}, we consider the inner $1~\rm{pc}$ of the Galaxy, 
where the values of the radiation energy density are $5\times10^4~\rm{eV/cm^3}$ at $kT=3.0~\rm{eV}$ (UV) and $kT=0.3~\rm{eV}$ (NIR), and $5\times 10^3~\rm{eV/cm^3}$ at $kT=6\times10^{-3}~\rm{eV}$ (FIR) \citep{dav}.
To have an equal energy density in the magnetic field, it is required that $B\simeq 500~\rm{\mu G}$, any lower value implies the dominance of IC (KN regime) losses in determining the cut-off in the electron spectrum. We choose $B_1=100~\rm{\mu G}$, a value of the magnetic field compression $\xi=4$, appropriate to Alfvenic turbulence, and a Bohm-type diffusion coefficient. For a completely randomised magnetic field, the compression factor would approach $\xi\simeq\sqrt{11}$.
As shown in the Fig. \ref{lossSNR}b, for a velocity of $3000~\rm{km/s}$, $B_1=100~\mu$G and $B_2=4B_1$, the cut-off energy is $E^*\simeq 10~\rm{TeV}$.

As discussed in Sect. \ref{Results}, depending on the age of the system, the particles in the low-energy part of the interval we consider may not have enough time to cool. Keeping the age of the historical SNR as a reference, which are known to be efficient electron accelerators, we performed our time-dependent calculation and present here the results for $\tau_{age}=100$ yr, 500 yr and finally, as a comparison, the steady state case.

At the shock location, steady state is reached quickly so that the three cases overlap perfectly.
The result is shown in Fig. \ref{electGC}a where we compare it with the one obtained for pure synchrotron losses, with the same values of the other parameters. The features are those pointed out in Sect. \ref{Results}: the cut-off region is broader and the decay is not as steep as for synchrotron cooled electrons.

In Fig. \ref{electGC}b we plot the spatially integrated spectra for the three different time scales chosen. The up and downstream contributions are also plotted for the steady state solution to show the nature of the substructure present around the cut-off energy. In the integrated distribution, the modification of the spectrum due to KN effects can be dramatic. In our example the hardening of the spectrum right below the cut-off causes a pile-up of more than one order of magnitude, compared to the solution for Thomson losses, which would correspond to a horizontal line in the plot. The shape of the spectrum around the cut-off is determined by losses, therefore the same as for the three curves. At lower energies, the point where losses set in depends on the time. For those energies where $\tau_{L}>\tau_{age}$, we obtain the uncooled spectrum $F(p)\propto p^{-4}$. We plot the steady state solution as a test to show that our method recovers the right solution $\propto p^{-5}$ at low energies where losses proceed in the Thomson regime.

Once we have the electron spectrum, we can calculate the photon distribution. Our results for IC are shown in Fig. \ref{radGC}a, where we plot the spectra for the different time scales we are considering. Each of the curves represents the sum of the contributions of the three components of the seed field (FIR, NIR, UV).
For the steady state solution, the overall radiation spectrum shows an almost flat profile below the cut-off due to the compensation of the ``double action'' of the Klein-Nishina effect. The other two curves harden in the low-energy part due to the slope of the electron spectrum at low energies. In addition, we see from Fig. \ref{lossSNR}b that the contribution of the three radiation fields dominate in different energy ranges. The higher the temperature of the black body, the lower the energy of the electrons required to enter the KN regime. For this reason, at early stages of the evolution (100 yr, dotted curve in Fig. \ref{radGC}a) when the low energy part of the electron spectrum is not yet contributing to the photon emission, the contribution of the two high temperature black bodies at 3500 K and 35000 K is almost zero and the total IC radiation is lower than at later times.

Fig. \ref{radGC}b shows the synchrotron emission. Here the situation is different: as expected the radiation spectrum reproduces the features of the electron one. A pronounced pile-up appears around the cut-off energy for all the three cases.

In this treatment we have assumed a value $B_1=100~\mu$G, reasonable for the Galactic centre region, although we should note that the magnetic field in this region is quite uncertain. The effect of a possible amplification of the magnetic field, according to \cite{ampl} and \cite{bellsat}, may lead to values of the magnetic field upstream of several hundreds of microGauss and more \citep{rxj_var}. Moreover, in the case of a modification of the shock due to the presence of accelerated particles, the total compression ratio can exceed the value of 4, therefore leading to an additional increase in the magnetic field value downstream. As a check, we performed our calculation for a compression ratio $R=7$ and a magnetic field compression $\xi=7$ and found that the effect of IC losses ceases to be significant, as compared to synchrotron cooling, for values of the upstream magnetic field exceeding 300 $\mu$G.


%

\subsection{Clusters of galaxies}
\label{Clusters of Galaxies}

\begin{figure}[]
\resizebox{\hsize}{!}{\includegraphics{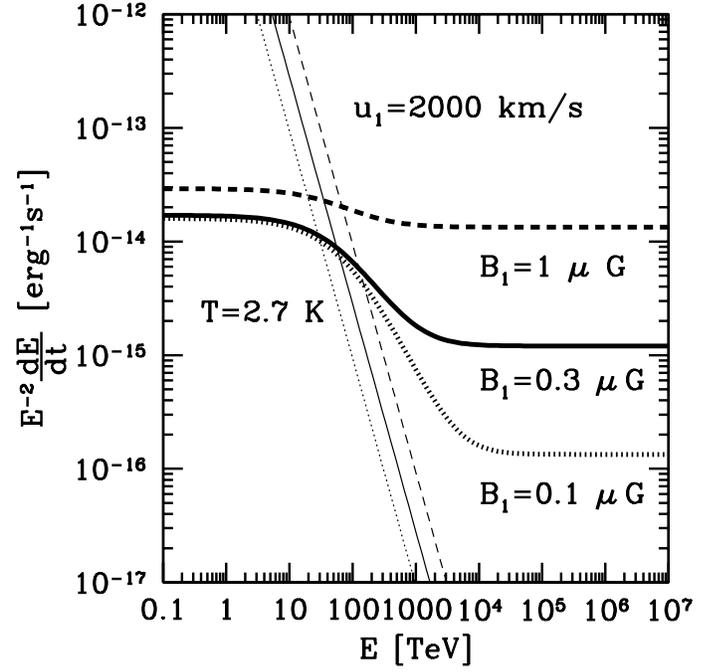}}
\caption{Thick lines: energy loss rates (synchrotron plus IC) for typical conditions in a galaxy cluster accretion shock. The background radiation field is provided by the CMB and the values of the magnetic field upstream are reported (downstream $B_2=4B_1$). Thin lines: acceleration rates for a shock velocity of $2000~\rm{km/s}$ and the line type refers to the corresponding value of $B_1$.}
\label{lossCl}
\end{figure}

\begin{figure}[]
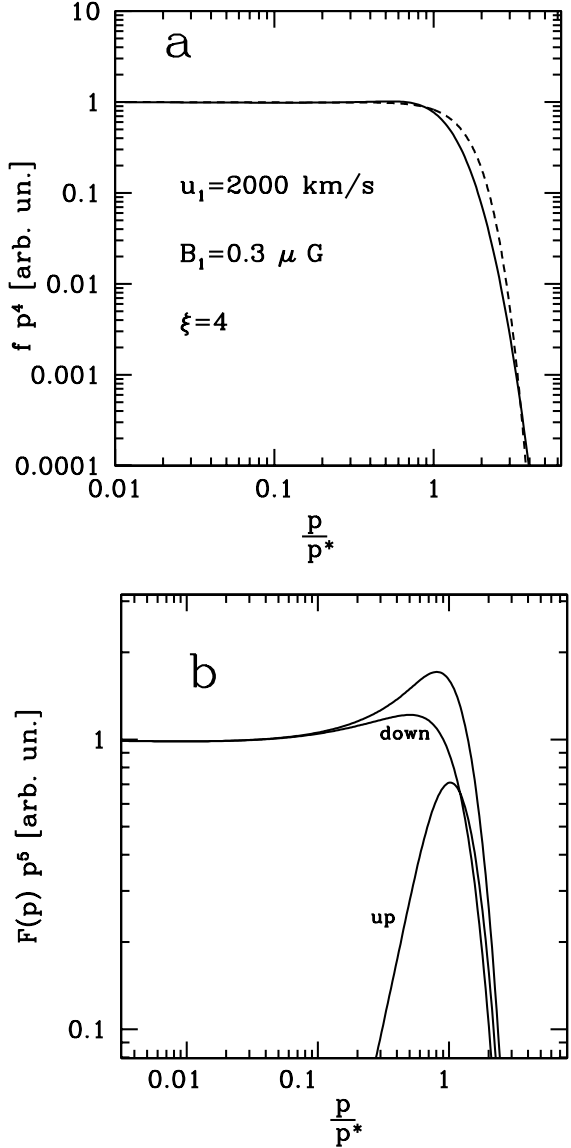

\includegraphics[width=0.4\textwidth]{9744fig13.epsi}\\
\\
\vspace{1cm}
\includegraphics[width=0.4\textwidth]{9744fig14.epsi}
\caption{\textbf{a)} Electron distribution at the shock front for a shock velocity of $2000~\rm{km/s}$ and a magnetic field $B_1=0.3~\mu$G in the two cases of intense inverse Compton losses (solid line) and dominant synchrotron cooling.
\textbf{b)} Integrated spectra for the IC dominated case. The upstream and downstream contributions are shown.}
\label{electCl}
\end{figure}

\begin{figure}[]
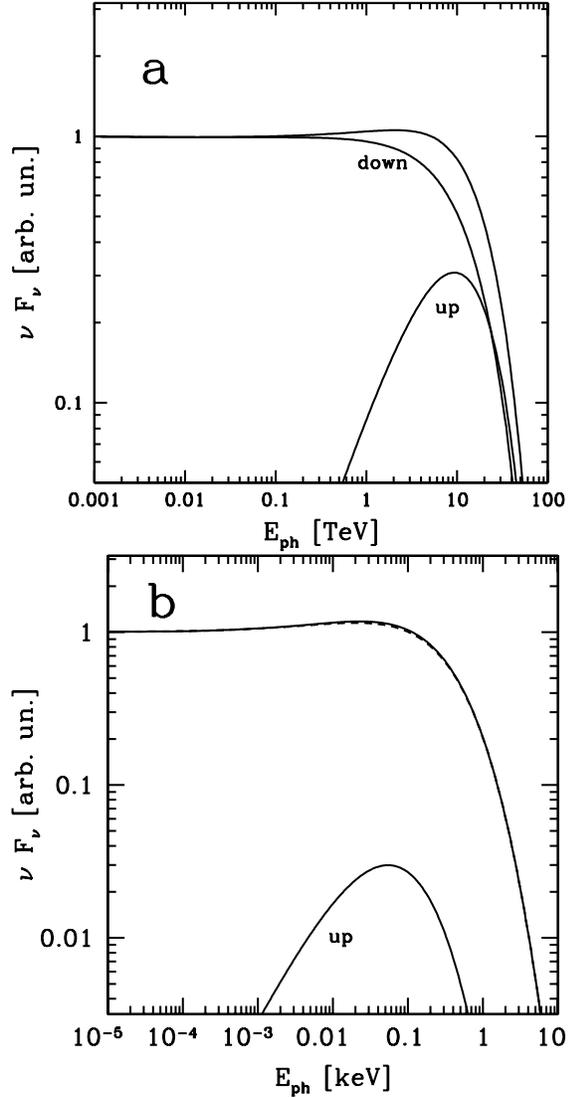

\includegraphics[width=0.4\textwidth]{9744fig15.epsi}
\\
\includegraphics[width=0.4\textwidth]{9744fig16.epsi}
\caption{\textbf{a)} Inverse Compton spectrum produced by upscattering of the CMB for the distributions in Fig \ref{electCl}b.
\textbf{b)} Synchrotron radiation emitted in the upstream region, downstream region, and their sum. The upstream contribution is negligible, so that the downstream curve (dashed) and the total one coincide.}
\label{radCl}
\end{figure}

Another example of a radiation-dominated environment where a shock can form is represented by clusters of galaxies.
Rich clusters of galaxies are the biggest virialised structures in the Universe, with typical sizes of a few Mpc and masses up to $10^{15} M_{\odot}$ or more (see \citealp{sarazinbook} for a review). 
An expanding shock wave, called the accretion shock, forms at the cluster boundary and carries the information of virialisation outward \citep{bertschinger}.
The infalling matter crosses the shock surface at a speed roughly comparable to the free fall velocity, namely,
$$
v_{s} \sim \sqrt{\frac{2 G M_{cl}}{R_{cl}}} \approx 2000 \left( \frac{M_{cl}}{10^{15} M_{\odot}} \right)^{1/2}  \left( \frac{R_{cl}}{3 Mpc} \right)^{-1/2} km/s,
$$
so that the shock velocities are comparable to the one found in supernova remnants.
Particle acceleration is expected to happen at accretion shocks (see \citealp{revste} for a review) with consequent emission of inverse Compton gamma rays, possibly detectable in the near future by GLAST or ground-based Cherenkov telescopes \citep{ste}.

The detection of a tenuous and diffuse synchrotron radio emission from about one third of the rich clusters of galaxies \citep{feretti,carilli} reveals the presence of a magnetic field of a few $\rm{\mu G}$ in the intracluster medium. The value of the magnetic field in the cluster outskirts, where the accretion shock propagates, is unknown, due to lack of radio measurements, but it is likely to be smaller or at most comparable to the one measured in the inner regions.
Thus, the magnetic-field energy density at the shock position is $\sim 0.025~\rm{eV/cm^3}$ for $1~\rm{\mu G}$ magnetic field.
This implies that, at the position of the shock, the main energy loss channel for relativistic electrons is inverse Compton scattering in the CMB photon field, characterised by an energy density of $\sim 0.25~\rm{eV/cm^3}$.
Optical and infrared radiation from cluster galaxies is totally negligible.

In Fig. \ref{lossCl} we plot the energy loss rates for a fiducial galaxy cluster. The thin lines represent the acceleration rates for a velocity $u_1=2000~\rm{km/s}$ and different values of the magnetic field. The thick lines are the total energy loss rates for inverse Compton scattering off the CMB plus synchrotron emission. 
Calculations are done for three values of the upstream magnetic field $B_1$ ($1~\rm{\mu G}$, $0.3~\rm{\mu G}$, $0.1~\rm{\mu G}$) and $B_2=4B_1$.

Below we consider the case of parameter values $B_1=0.3~\mu$G and $u_1=2000~\rm{km/s}$, which are reasonable for an accretion shock around a rich cluster \citep{revste}. In this case, the cut-off energy falls into the KN-dominated part of the total loss rate, at about $60~\rm{TeV}$ (see Fig. \ref{lossCl}). At the cut-off energy the loss time scale is of the order of $4\times10^4$ yr. The evolution times of large-scale structures are several Gyr, meaning that electrons cool comparatively rapidly. Therefore the steady state is always reached.

In Fig. \ref{electCl}a we plot the electron spectrum at the shock, compared to the one expected for pure synchrotron losses. The modification due to the dominant IC losses has the same characteristics as noticed before: a wider and less sharp cut-off.
If we look at the effect this produces on the total spatially integrated particle spectrum, we find that a small pile-up is still present around the cut-off energy (Fig. \ref{electCl}b). This feature is less pronounced in the present case because the energy losses right below the cut-off energy fall in the Thomson regime, as we can see from Fig. \ref{lossCl}. This process does not happen deep enough in the Klein-Nishina regime to produce a strong feature.

The photon spectra are affected in a similar way. In Figure \ref{radCl}a we plot the upscattered spectrum of the CMB. We plot the two contributions to the total spectrum from the upstream and downstream regions, the latter dominating the total spectrum. 
Fig. \ref{radCl}b shows the synchrotron spectrum where the small pile-up feature of the electron spectrum is reproduced.

\subsection{Microquasars}
\label{Microquasars}

\begin{figure}[]
\resizebox{\hsize}{!}{\includegraphics{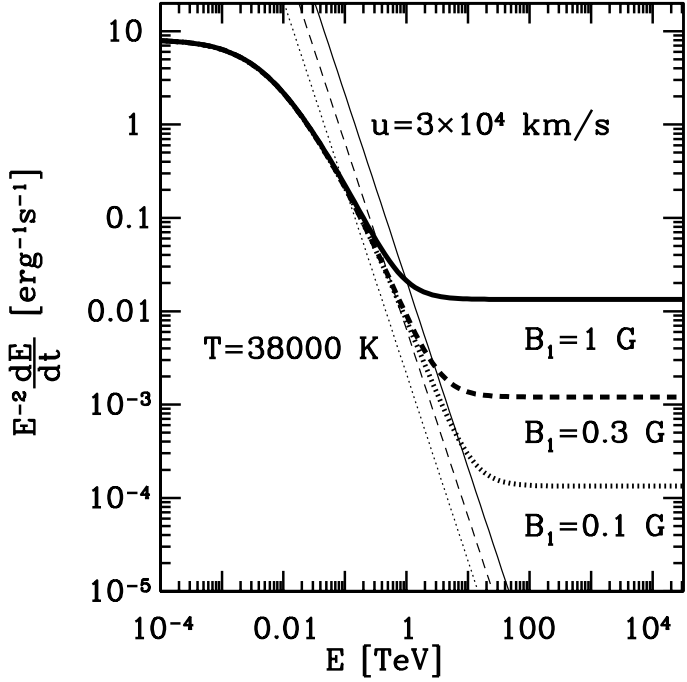}}
\caption{Total loss rates (thick lines) due to IC upscattering of the star photon field at $kT\simeq 3.3~\rm{eV}$ plus synchrotron emission for the values of $B_1$ reported and $B_2=4B_1$, feasible for a Microquasar. The shock velocity is fixed at $30000~\rm{km/s}$ and the thin lines correspond to the acceleration rates connected to the values of the magnetic field for which we plot the loss rates.}
\label{lossMQ}
\end{figure}

\begin{figure}[]
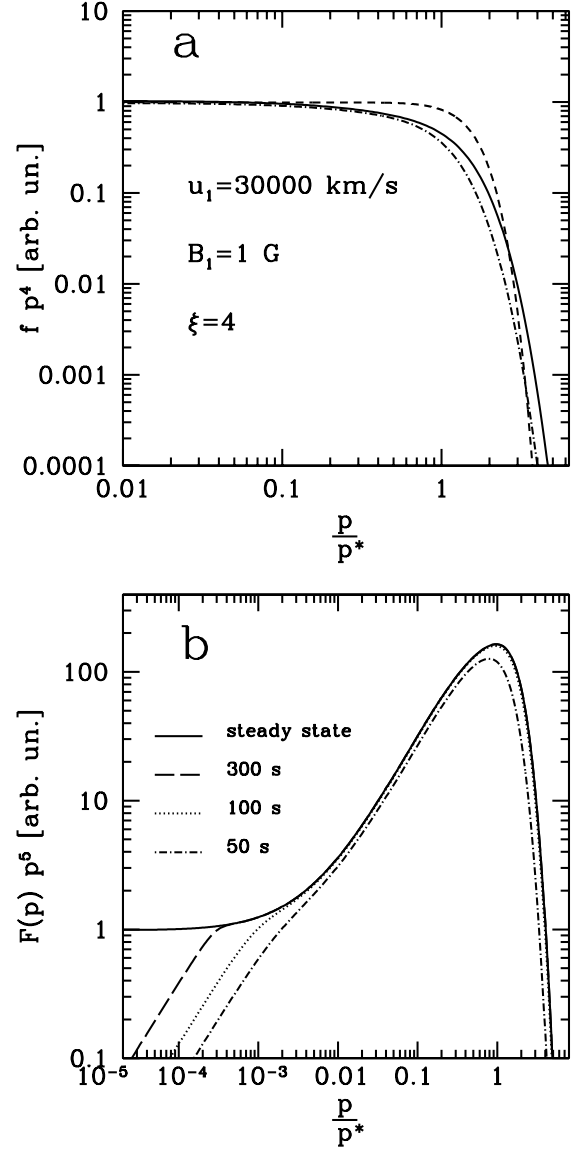

\includegraphics[width=0.4\textwidth]{9744fig18.epsi}\\
\\
\includegraphics[width=0.4\textwidth]{9744fig19.epsi}
\caption{\textbf{a)} Electron spectrum at the shock for a shock velocity $u_1=30000~\rm{km/s}$. The value of the magnetic field is $B_1=1~\rm{G}$ and $B_2=4B_1$. The solid line represents the case of dominant IC losses for an acceleration time $\geq 100$s, the dot-dashed one is for $t_{acc}=50s$, while the dashed one is obtained with pure synchrotron cooling in steady state.
\textbf{b)} Integrated electron spectra for the time scales reported.}
\label{electMQ}
\end{figure}

\begin{figure}[]
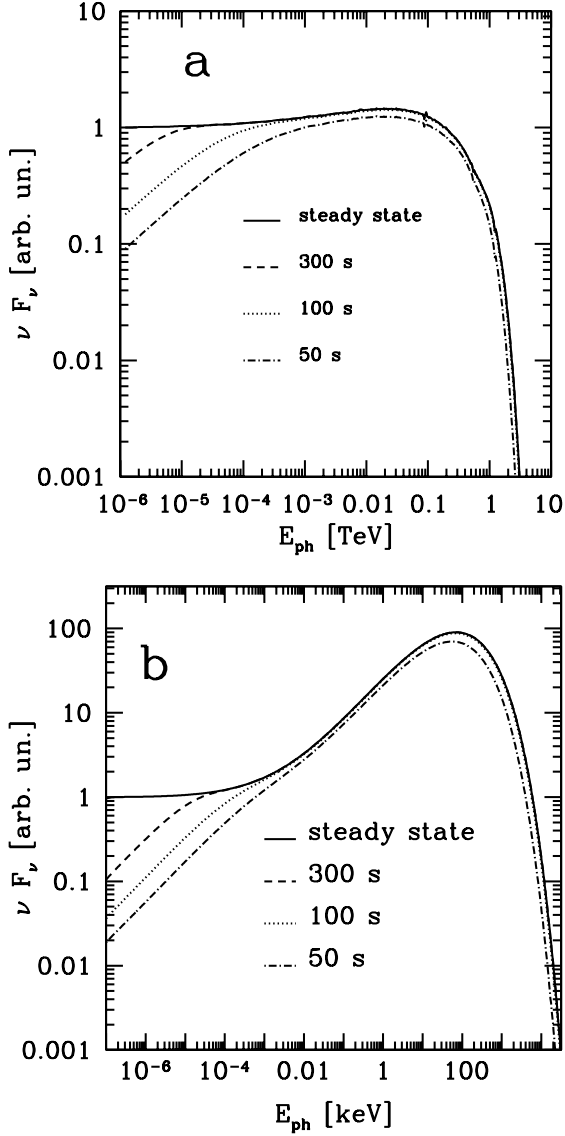

\includegraphics[width=0.4\textwidth]{9744fig20.epsi}\\
\\
\includegraphics[width=0.4\textwidth]{9744fig21.epsi}
\caption{Photon spectra produced by the electron distributions in Fig. \ref{electMQ}b. \textbf{a}) IC spectra for a diluted black body of temperature $T= 3.8\times10^4~\rm{K}$.
\textbf{b)} Synchrotron spectra.}
\label{radMQ}
\end{figure}
Another class of objects of interest are the microquasars. These are galactic binary systems composed of a regular O/B star being accreted onto a compact object (neutron star or black hole) that presents a jet.
The nature of these objects is very complex and is not yet fully understood.
Nevertheless the basic ingredients present in the system are a very intense stellar radiation field; e.g., for LS 5039, the luminosity of the star is $L_*\simeq7\times10^{38}~\rm{erg/s}$ implying a radiation energy density up to $U_{rad}\simeq 1000~\rm{erg/cm^3}$, depending on the location, at a temperature of $T\simeq 3.8\times10^4~\rm{K}$ \citep{casares}, and a jet that is thought to be the site of particle acceleration. If blobs of plasma are emitted in the jet at slightly different velocity, internal sub-relativistic shocks can form, where electron acceleration can take place. The value of the magnetic field is not known, but has a reasonable range between $0.01$ and $1~\rm{G}$.

For our calculation we refer to the model used in \citet{mitya}, where the jet from the compact object is assumed to have an axis perpendicular to the orbital plane, and we put the acceleration zone at a distance $Z_0=2\times10^{12}~\rm{cm}$ from the jet base (corresponding to a distance equal to the mean orbital radius). At such a distance, the radiation energy density results:

\begin{equation}
\frac{L_*}{4\pi c(R_{orb}^2+Z_0^2)}\simeq 230 \frac{erg}{cm^3}.
\end{equation}

In Fig. \ref{lossMQ} we plot the total energy loss rates due to IC plus synchrotron losses and the acceleration rates for a shock velocity of $30000~\rm{km/s}$, for three values of the upstream magnetic field $B_1=1~\rm{G}$, $0.3~\rm{G}$, $0.1~\rm{G}$, and a downstream magnetic field $B_2=4B_1$. We chose $u_1=30000~\rm{km/s}$, corresponding to 10\% of the speed of light, in order to remain consistent with the assumption of nonrelativistic shock.

In this case the radiation energy density is considerably higher than the magnetic one, even for a field $B_1=1~\rm{G}$. This is the value we considered in our calculation in order to obtain an efficient acceleration of electrons up to TeV energies. Moreover, from a theoretical point of view, this choice allows us to explore a different region of the KN losses, compared with the previous two cases.

The IC cooling time at the cut-off energy for the configuration described is $\tau_L\simeq 40$ s. As a rough estimate, we can say that the dynamical time of the shock is $\tau_{age}\sim Z_0/u_1\simeq 300~\textrm{s}$. We therefore report on the results of our calculation for a time of 50 s, 100 s, and 300 s. We also show the steady state solution for comparison.
In Fig. \ref{electMQ} we plot the electron spectrum at the shock and integrated over space. The difference in the shape of the spectrum around the cut-off energy between the expected spectrum for pure synchrotron cooling and the one obtained for dominant IC losses in panel \textbf{a} is very pronounced. We see that, when the age of the system is comparable to the loss time scale at the cut-off energy (50 s), the steady state is not reached even at the high energy end of the spectrum (Fig. \ref{electMQ}, both panels). In the cases corresponding to 100 s and 300 s, on the other hand, the spectrum around the cut-off reaches its steady state. Only the low energy part of the integrated spectrum presents the characteristic uncooled slope (Fig. \ref{electMQ}b).

In this case the predominance of the IC is such that the hardening of the integrated spectrum is very pronounced, the pile-up spanning two orders of magnitude.

As in the previous cases, the IC photon spectrum does not show prominent features (Fig. \ref{radMQ}a), while the synchrotron one reproduces the electron pile-up (Fig. \ref{radMQ}b).

\section{Conclusions}
\label{Conclusions}

We have addressed the problem of particle acceleration at nonrelativistic shocks in the presence of strong energy losses, specifically in the never before treated case of electrons undergoing severe inverse Compton losses in the Klein-Nishina regime.

We numerically solved the complete transport equation for the accelerated particles and calculated the particle distribution function $f(x,p)$ and the resulting radiation spectra.
Our approach provides a general tool for what concerns the species of particles accelerated (electrons and protons), the type of losses and the form of the diffusion coefficient. We focused on electrons undergoing synchrotron and inverse Compton losses, where the radiation energy density is higher than the magnetic one.
We assumed Bohm diffusion: $D(p)=pc^2/(3eB)$.

For high-energy electrons, with energy $E\ge (mc^2)^2/\epsilon_{ph}$, the inverse Compton scattering enters the Klein-Nishina regime where the cross-section decreases with energy. We evaluated the particle spectrum at the shock and found that, compared to the case of Thomson losses, the KN effect results in a broadening of the cut-off region and a hardening of the spectrum around the cut-off energy. This effect is most evident in the spatially integrated spectrum, which exhibits a pronounced pile-up feature just below the cut-off energy. Such a feature can be significant, with an enhancement of the spectrum up to a few orders of magnitude, depending on the ratio of energy densities $U_{rad}/U_{mag}$, the magnetic field strength, and the shock velocity.

Once the electron spectrum is obtained, we can calculate the resulting spectra of the radiation, both synchrotron and inverse Compton. The hardening in the electron spectrum is barely visible in the IC spectrum since the effect of the KN cross-section on the scattering process itself compensates the opposite action on the electron spectrum. The pile-up feature is reproduced by the synchrotron distribution, which no longer follows a simple power law below the cut-off, as in the case of dominant Thomson losses.

We demonstrated the importance of this effect in three examples of astrophysical objects in which the energy density of the radiation field dominates the magnetic field energy density, namely a supernova remnant located in the Galactic centre region, the accretion shocks of clusters of galaxies and the internal shocks in the jet of a microquasar, and discussed the possible consequences of our results on the emission of these objects.

\begin{acknowledgements} 
We especially thank V. Zirakashvili for all the useful advice and suggestions. We would also like to acknowledge V. Bosch-Ramon, D. Khangulyan, J. G. Kirk, B. Reville, and A. M. Taylor for fruitful discussions. We thank the referee, P. Blasi, for useful comments. GV acknowledges support from the International Max-Planck Research School (IMPRS) Heidelberg. SG acknowledges the support of the European Community under a Marie Curie Intra-European fellowship.
\end{acknowledgements}

\bibliographystyle{aa}

{}

\end{document}